\documentclass[a4paper,10pt]{article}

\usepackage{amssymb}

\newtheorem{definition}{Definition}[section]
\newtheorem{lemma}[definition]{Lemma}
\newtheorem{proposition}[definition]{Proposition}
\newtheorem{theorem}[definition]{Theorem}
\newtheorem{corollary}[definition]{Corollary}

\newenvironment{proof*}{\smallskip\par\noindent\emph{Proof. }
 \ignorespaces}{\hfill$\Box$\smallskip\par\ignorespaces}
\newenvironment{proofsketch*}{\smallskip\par\noindent
 \emph{Sketch of proof. }\ignorespaces}
 {\hfill$\oslash$\smallskip\par\ignorespaces}
\newenvironment{example}{\smallskip\par\noindent
\textbf{Example:\ }}{\hfill$\oslash$\smallskip\par\ignorespaces}

\newcommand{\map}[3]{\ensuremath{#1\!:\!#2\!\rightarrow\!#3}}
\newcommand{\N}{\ensuremath{\mathbb{N}}}
\newcommand{\R}{\ensuremath{\mathbb{R}}}
\newcommand{\C}{\ensuremath{\mathbb{C}}}
\newcommand{\Alg}[1]{\ensuremath{\mathcal{#1}}}
\newcommand{\Test}{\ensuremath{C^{\infty}}}

\title{Equivalence of the (generalised) Hadamard and microlocal spectrum
condition for (generalised) free fields in curved spacetime}
\author{Ko Sanders\thanks{E-mail:
jacobus.sanders@theorie.physik.uni-goe.de}\\
Institute of Theoretical Physics\\
University of G\"ottingen,\\
Friedrich-Hund-Platz 1, D-37077 G\"ottingen\\
and\\
Courant Research Centre\\
''Higher Order Structures in Mathematics'',\\
University of G\"ottingen}
\date{3 February 2009}

\begin{document}

\maketitle

\begin{abstract}
We prove that the singularity structure of all $n$-point distributions
of a state of a generalised real free scalar field in curved spacetime can
be estimated if the two-point distribution is of Hadamard form. In particular
this applies to the real free scalar field and the result has applications in
perturbative quantum field theory, showing that the class of all Hadamard
states is the state space of interest. In our proof we assume that the field
is a generalised free field, i.e.~that it satisfies scalar (c-number)
commutation relations, but it need not satisfy an equation of motion. The
same argument also works for anti-commutation relations and it can be
generalised to vector-valued fields. To indicate the strengths and limitations
of our assumption we also prove the analogues of a theorem by Borchers and
Zimmermann on the self-adjointness of field operators and of a very weak form
of the Jost-Schroer theorem. The original proofs of these results in the
Wightman framework make use of analytic continuation arguments. In our case no
analyticity is assumed, but to some extent the scalar commutation relations can
take its place.
\end{abstract}

\section{Introduction}

The study of quantum field theories in curved spacetime is simplified
considerably by the use of techniques from microlocal analysis to study
the singularities of $n$-point distributions. Ever since Radzikowski
\cite{Radzikowski} has shown that Hadamard states of the real free
scalar field can be characterised by the wave front set of their
two-point distributions, these techniques have been on the increase as a
suitable replacement of the Fourier transform in Minkowski spacetime. This
enabled  \cite{Brunetti+1} to introduce a microlocal spectrum condition
($\mu$SC) for general real scalar fields which is a smoothly covariant
condition that generalises Wightman's spectrum condition.

The generalisation is only possible at a price: whereas the
$n$-point distributions of a Wightman field are the boundary values of
analytic functions, this is no longer so in curved spacetimes. In fact,
a generic curved spacetime cannot be expected to be analytic at all, so
all arguments involving analytic continuation have to be reexamined in
the context of quantum field theory in curved spacetime. In
\cite{Strohmaier+} an analytic microlocal spectrum condition was
introduced on analytic spacetimes in order to provide an amount of
analyticity analogous to the Wightman case, but the requirement that the
metric be analytic in some analytic structure on the manifold, although
technically advantageous, seems to be unphysically restrictive.

In this work we will not require any analyticity, but instead we consider
a real scalar field which satisfies scalar (i.e.~c-number) commutation
relations\footnote{Our results also work for anti-commutation relations
and for vector-valued fields.}. These fields, which include the real free
scalar field, will be called generalised free fields, following the
terminology of the Wightman framework in Minkowski spacetime (see
e.g.~\cite{Jost}), although in curved spacetime not much seems to be known
about them. As our main result we will prove that an estimate on the
singularities of the two-point distribution (''generalised Hadamard
condition'') implies estimates on the singularities of all $n$-point
distributions. In particular, all truncated $n$-point distributions with
$n\not=2$ will be shown to be smooth and consequently the state will
satisfy the $\mu$SC. An easy application is that the class of generalised
Hadamard states is closed under operations from the algebra of observables.
Moreover, all Hadamard states of a free field can be extended to the
extended algebra of Wick polynomials and time-ordered products as
constructed by Hollands and Wald \cite{Hollands+1,Hollands+2}.

After that we will investigate the strength of our assumption by proving the
analogues of a result by Borchers and Zimmermann on the self-adjointness of
field operators and a very weak version of the Jost-Schroer theorem. In both
cases the original proofs rely on analytic continuation arguments, but in our
case no analyticity is assumed. Instead, the commutation relations take the
place of analyticity to a certain extent, but not fully. Indeed, we have
weakened the statement of the Jost-Schroer theorem to compensate for the
change in assumptions.

The organisation of our paper is as follows: we first establish our
notation for quantum field theory in curved spacetime in section
\ref{sec_QFT}. There we also present the microlocal spectrum condition, the
(generalised) Hadamard condition and the truncated $n$-point distributions
and we collect some results concerning the singularities of the two-point
distribution. In section \ref{sec_GFF} we introduce the commutation relations
and give two equivalent characterisations of generalised free fields.
Section \ref{sec_npoint} contains our main results concerning the
singularity structure of higher $n$-point distributions and truncated
$n$-point distributions, as well as a result on the comparison of $n$-point
distributions of different states. In section \ref{sec_2thm} we discuss the
generalisations of the result by Borchers and Zimmermann and the Jost-Schroer
theorem. We conclude with some easy applications and an outlook in section
\ref{sec_appl}. For an introduction to microlocal analysis we refer to
chapter 8 of \cite{Hoermander}.

\section{Real scalar quantum fields and the microlocal spectrum
condition}\label{sec_QFT}

Let $M=(\mathcal{M},g)$ be a spacetime, i.e.~$\mathcal{M}$ is a
smooth, connected manifold of dimension $D\ge 2$ with the smooth
Lorentzian metric $g$, where we use the signature convention
$+-\ldots -$. We let $V\subset TM$ denote the set of all causal
tangent vectors (including $0$-vectors) and we let $V^*\subset T^*M$
be its dual, i.e.~the image of $V$ under the identification of $TM$
with $T^*M$ via the metric. We assume that $M$ is time-oriented, so
we can define the future and past causal cones $V^{\pm}\subset TM$
and their duals, $V^{*\pm}\subset T^*M$. We use $\mathcal{Z}$ to
denote the zero section of a vector bundle (it will always be clear
from the context which vector bundle is meant).

A real scalar quantum field on the spacetime $M$ can be described
using the Borchers-Uhlmann algebra. Here we adopt the convention that
the space $M^0$ consists of a single point, so that $\Test_0(M^0)=\C$.
\begin{definition}
The (scalar) \emph{Borchers-Uhlmann} algebra on the spacetime $M$ is
defined to be the topological $^*$-algebra
$\Alg{U}_M:=\oplus_{n=0}^{\infty} \Test_0(M^{\times n})$, where we
allow only finite direct sums and where
\begin{enumerate}
\item the product is determined by the linear extension of\\
$f(x_{n+m},\ldots,x_{n+1})g(x_n,\ldots,x_1):=
(f\otimes g)(x_{n+m},\ldots,x_1)$,
\item the $^*$-operation is determined by anti-linear extension of\\
$f^*(x_n,\ldots,x_1):=\overline{f(x_1,\ldots,x_n)}$,
\item as a topological space $\Alg{U}_M$ is the strict inductive limit
\[
\Alg{U}_M=\cup_{N=0}^{\infty}\oplus_{n=0}^N\Test_0(K_N^{\times n}),
\]
where $K_N$ is an exhausting (and increasing) sequence of compact
subsets of $\mathcal{M}$ and each $\Test_0(K_N^{\times n})$ is given
the test-function topology (cf. \cite{Schaefer} theorem 2.6.4).
\end{enumerate}

A \emph{state} on the Borchers-Uhlmann algebra is a normalised
continuous positive linear map $\map{\omega}{\Alg{U}_M}{\C}$.
\end{definition}
The topology of $\Alg{U}_M$ is such that $f_j=\oplus_n f_j^{(n)}$
converges to $f=\oplus_n f^{(n)}$ if and only if for all $n$ we have
$f_j^{(n)}\rightarrow f^{(n)}$ in $\Test_0(M^{\times n})$ and 
all $f_j^{(n)}$ vanish if $n\ge N$ for some $N>0$. A state
therefore consists of a sequence of $n$-point distributions,
$\omega=\left\{\omega_n\right\}_{n=0}^{\infty}$, where $\omega_n$ is a
distribution on $M^{\times n}$. The algebra $\Alg{U}_M$ has the unit
$I=1\oplus 0\oplus 0\ldots$ and the normalisation of the state $\omega$
means that $\omega(I)=\omega_0=1$. Given a state one can construct the
GNS-representation $\pi_{\omega}$ on a Hilbert space $\mathcal{H}_{\omega}$
with a dense domain $\mathcal{D}_{\omega}$ that contains a vector
$\Omega_{\omega}$ such that:
$\mathcal{D}_{\omega}=\pi_{\omega}(\Alg{U}_M)\Omega_{\omega}$ and
$\omega(A)=\langle\Omega_{\omega},\pi_{\omega}(A)\Omega_{\omega}\rangle$
for each $A\in\Alg{U}_M$. The GNS-quadruple
$(\pi_{\omega},\mathcal{H}_{\omega},\mathcal{D}_{\omega},\Omega_{\omega})$
is ths unique quadruple with these properties, up to unitary equivalence.

Instead of the $n$-point distributions one often considers the truncated
$n$-point distributions of a state $\omega$, which we will now define.
For $n\ge 1$ we let $\mathcal{P}_n$ denote the set of all partitions
of the set $\left\{1,\ldots,n\right\}$ into pairwise disjoint subsets,
which are ordered from low to high. If $r$ is an ordered set in the
partition $P\in\mathcal{P}_n$ we write $r\in P$ and we denote the
elements of $r$ by $r(1)<\ldots<r(|r|)$, where $|r|$ is the number of
elements in $r$. The truncated $n$-point distributions
$\omega^T_n$, $n\ge 1$, of a state $\omega$ are defined implicitly in
terms of the $n$-point distributions $\omega_n$ by:
\begin{equation}\label{eqn_deftrunc}
\omega_n(x_n,\ldots,x_1)=\sum_{P\in\mathcal{P}_n}\prod_{r\in P}
\omega^T_{|r|}(x_{r(|r|)},\ldots,x_{r(1)}).
\end{equation}
Note that this equation can be solved iteratively for $\omega^T_n$
order by order.
\begin{definition}\label{def_qf}
A state $\omega$ is called \emph{quasi-free} if and only if
$\omega_n^T\equiv 0$ for all $n\not=2$.
\end{definition}

We will denote by $\Phi$ the canonical injection
$\Test_0(M)\subset\Alg{U}_M$, which sends $f$ to
\[
\Phi(f):=0\oplus f\oplus 0\oplus\ldots .
\]
The map $\Phi$ is a distribution with values in $\Alg{U}_M$ and it
represents the real scalar quantum field. In the GNS-representation
the field is representation by $\Phi^{\omega}(f):=\pi_{\omega}(\Phi(f))$.
For our current purposes it is convenient not to impose commutation
relations, causality or an equation of motion on the field $\Phi$, but
to let them be dictated for $\Phi^{\omega}$ by the state. This will be
done in section \ref{sec_GFF}.

We now give an equivalent reformulation of the microlocal spectrum
condition due to \cite{Brunetti+1}, starting with the introduction
of some terminology.
\begin{definition}
We let $\mathcal{G}_n$ denote the set of all graphs with $n$ vertices
and finitely many edges. An \emph{immersion} of a graph
$G\in\mathcal{G}_n$ into the spacetime $M$ consists of an assignment
of
\begin{itemize}
\item a point $x(i)\in M$ to each vertex $\nu_i$ of $G$,
\item a piecewise smooth curve $\gamma_r$ between $x(i)$ and $x(j)$
to every edge $e_r$ of $G$ that connects $\nu_i$ and $\nu_j$,
\item a causal, future pointing covector field $\xi_r$ on $\gamma_r$ to
each $e_r$, so that $\xi_r$ is covariantly constant, $\nabla \xi_r=0$,
along $\gamma_r$.
\end{itemize}

An immersion of a graph $G\in\mathcal{G}_n$ into the spacetime $M$ is
called \emph{causal}, resp.~\emph{light-like}, iff the curves
$\gamma_r$ are causal, resp.~light-like.

We say that a point
$(x_n,k_n;\ldots;x_1,k_1)\in T^*M^n\setminus\mathcal{Z}$
is \emph{instantiated} by an immersion of a graph $G\in\mathcal{G}_n$
if and only if for each $i=1,\ldots,n$ the immersion sends the vertex
$\nu_i$ to $x_i$ and
\[
k_i=\sum_{e_r \mathrm{\ between\ } i
\mathrm{\ and\ } j>i}\xi_r(x_i)-\sum_{e_r \mathrm{\ between\ } j<i
\mathrm{\ and\ } i}\xi_r(x_i).
\]
\end{definition}
Recall that $\mathcal{Z}$ denotes the zero section of a vector bundle.
The covector field $\xi_r$ is to be thought of as a singularity,
propagating along the curve $\gamma_r$ from $x(i)$ to $x(j)$. The
following sets describe the singularities that we allow the
$n$-point distributions to have:
\begin{eqnarray}\label{def_cone}
\Gamma_n&:=&\left\{(x_n,k_n;\ldots;x_1,k_1)\in T^*M^n\setminus\mathcal{Z}
|\ \exists G\in\mathcal{G}_n\mathrm{\ and\ an\ immersion\ of\ }G
\right.\nonumber\\
&&\left.\mathrm{\ into\ } M \mathrm{\ which\ instantiates\ the\ point\ }
(x_n,k_n;\ldots;x_1,k_1)\right\}.
\end{eqnarray}
The sets $\Gamma_n^c$, resp.~$\Gamma_n^{ll}$, are defined similarly, but
using only causal, resp.~light-like, immersions of graphs. In general we
will write $\Gamma_n^{\bullet}$, where $^{\bullet}$ denotes either no
superscript, or $c$ or $ll$.

\begin{definition}
A state $\omega$ 
satisfies the \emph{microlocal spectrum condition}
($\mu$SC) with smooth, resp.~causal, resp.~light-like immersions, iff
for all $n\in\N$ we have $WF(\omega_n)\subset\Gamma_n^{\bullet}$,
where $^{\bullet}$ denotes no superscript, resp.~$c$, resp.~$ll$.

If $M$ is an analytic spacetime then $\omega$ satisfies the
\emph{analytic microlocal spectrum condition} (A$\mu$SC) with smooth,
resp.~causal, resp.~light-like immersions, iff for all $n\in\N$ we have
$WF_A(\omega_n)\subset\Gamma_n^{\bullet}$, where $^{\bullet}$ denotes no
superscript, resp.~$c$, resp.~$ll$.
\end{definition}
The usefulness of these restrictions on the singularities of the
$n$-point distributions derives largely from the properties of the
sets $\Gamma_n^{\bullet}$:
\begin{proposition}\label{propGamma}
The sets $\Gamma_n^{\bullet}$, with a fixed choice for the superscript
$^{\bullet}$, have the following properties:
\begin{enumerate}
\item each
$\Gamma_n^{\bullet}\subset T^*M^{\times n}\setminus\mathcal{Z}$
is a convex cone,
\item $\Gamma_n^{\bullet}\cap-\Gamma_n^{\bullet}=\emptyset$,
\item $\pi((\Gamma_{n_1}^{\bullet}\cup\mathcal{Z})\times\ldots\times
(\Gamma_{n_m}^{\bullet}\cup\mathcal{Z}))\subset\Gamma_{n_1+\ldots+n_m}^{\bullet}
\cup\mathcal{Z}$, where $\pi$ is a permutation
acting on the indices such that
$\pi(1)<\pi(2)<\ldots<\pi(n_1)$, $\pi(n_1+1)<\ldots<\pi(n_1+n_2)$,
\ldots, $\pi(n_1+\ldots+n_{m-1}+1)<\ldots<\pi(n_1+\ldots+n_m)$,
\item $(x_1,k_1;\ldots;x_n,k_n)\in-\Gamma_n^{\bullet}$ iff
$(x_n,k_n;\ldots;x_1,k_1)\in\Gamma_n^{\bullet}$.
\end{enumerate}
\end{proposition}
\begin{proof*}
We refer to \cite{Brunetti+1} lemma 4.2 for a proof of the first
property. The second property follows from the first and the third property
follows immediately from the definitions, using the unions of disconnected
instantiating graphs (cf.~\cite{Brunetti+1} proposition 4.3). The
fourth property follows directly from the definitions.
\end{proof*}

\begin{lemma}\label{lem_mSCwT}
A state $\omega$ satisfies the \emph{microlocal spectrum condition} ($\mu$SC)
with smooth, resp.~causal, resp.~light-like immersions iff
$WF(\omega^T_n)\subset\Gamma_n^{\bullet}$ for all $n\in\N$, where
$^{\bullet}$ denotes no superscript, resp.~$c$, resp.~$ll$. The same result
holds in the analytic case.
\end{lemma}
\begin{proof*}
We prove by induction on $n\in\N$ that $WF(\omega^T_n)\subset\Gamma_n^{\bullet}$
if and only if $WF(\omega_n)\subset\Gamma_n^{\bullet}$. For
$n=1$ this holds because $\omega^T_1=\omega_1$. Now assume that the claim holds
for $i=1,\ldots,n-1$ for some $n\ge 2$. From equation (\ref{eqn_deftrunc}) we
see that $\omega_n-\omega^T_n$ can be expressed as a sum, whose wave front set
is contained in $\Gamma_n^{\bullet}$ by items 1 and 3 in proposition
\ref{propGamma}. Using item 1 of this proposition once more we see that
$WF(\omega_n)\subset\Gamma_n^{\bullet}$ if and only if 
$WF(\omega^T_n)\subset\Gamma_n^{\bullet}$. The argument is purely
combinatorical, so it remains true in the analytic case.
\end{proof*}

A much weaker condition than the microlocal spectrum condition is the Hadamard
condition. This condition only places a restriction on the singularities of the
two-point distribution so as to enable the renormalisation of the
stress-energy-momentum tensor of a free field (see \cite{Wald}). Because our field
need not be free we will consider the following immediate generalisation of the
Hadamard condition:
\begin{definition}\label{def_genHad}
A state $\omega$ on the Borchers-Uhlmann algebra $\Alg{U}_M$ of the
spacetime $M$ is called a \emph{generalised Hadamard state} iff
$WF(\omega_2)\subset\Gamma_2$.
\end{definition}
Note that a (generalised) Hadamard state need not be quasi-free. We will show
in section \ref{sec_GFF} that the generalised Hadamard condition reduces to
the Hadamard condition in the case of free fields.

To complete this section we will now collect some small but useful results
on the relation between the generalised Hadamard condition and the two-point
distribution. For this purpose we define the symmetric and anti-symmetric part
of the two-point distribution by
\begin{equation}\label{eqn_2pm}
\omega_{2\pm}(x_2,x_1):=\frac{1}{2}(\omega_2(x_2,x_1)\pm\omega_2(x_1,x_2)). 
\end{equation}
The idea of the following proof is taken from proposition 6.1 in
\cite{Strohmaier+}.
\begin{proposition}\label{prop2point}
If $\omega$ is a generalised Hadamard state, then we have:
\begin{itemize}
\item $(x_2,k_2;x_1,k_1)\in WF(\omega_{2\pm})$ iff
$(x_1,k_1;x_2,k_2)\in WF(\omega_{2\pm})$ iff\\
$(x_2,-k_2;x_1,-k_1)\in WF(\omega_{2\pm})$,
\item $WF(\omega_{2+})= WF(\omega_{2-})\subset\Gamma_2\cup-\Gamma_2$,
\item $WF(\omega_2)=WF(\omega_{2-})\cap\Gamma_2$.
\end{itemize}
\end{proposition}
\begin{proof*}
The positivity of $\omega$ implies that
$\overline{\omega_2}(x_2,x_1)=\omega_2(x_1,x_2)$ and hence that
$\overline{\omega_{2\pm}}(x_2,x_1)=\omega_{2\pm}(x_1,x_2)=
\pm\omega_{2\pm}(x_2,x_1)$, from which the first property follows. That
$WF(\omega_{2\pm})\subset\Gamma_2\cup-\Gamma_2$ for a generalised
Hadamard state is clear from the definition. Now suppose that
$(x_2,k_2;x_1,k_1)\in WF(\omega_{2\pm})$. Then we can distinguish two
cases, namely either $(x_2,k_2;x_1,k_1)\not\in WF(\omega_2)$ or
$(x_2,-k_2;x_1,-k_1)\not\in WF(\omega_2)$ by equation (\ref{eqn_2pm})
and statement 2 of proposition \ref{propGamma}. Using
$\omega_{2\pm}=\omega_2-\omega_{2\mp}$ and the properties under the first
item we find that either
$(x_2,k_2;x_1,k_1)\in WF(\omega_{2\mp})$ or
$(x_2,-k_2;x_1,-k_1)\in WF(\omega_{2\mp})$, as the case may be, and hence
that $(x_2,k_2;x_1,k_1)\in WF(\omega_{2\mp})$. Thus
$WF(\omega_{2+})\subset WF(\omega_{2-})$ and the opposite inclusion
can be proved in the same way. For the last item we use again the
definition $2\omega_{2-}=\omega_2-\tilde{\omega}_2$, where
$\tilde{\omega}_2(x_2,x_1):=\omega_2(x_1,x_2)$. By the assumption on
$\omega_2$ we have $WF(\omega_2)\cap WF(\tilde{\omega}_2)=\emptyset$.
Hence we deduce: $WF(\omega_2)\subset WF(\omega_{2-})$,
$WF(\tilde{\omega}_2)\subset WF(\omega_{2-})$ and
$WF(\omega_{2-})\subset WF(\omega_2)\cup WF(\tilde{\omega}_2)$, from
which it follows that
$WF(\omega_{2-})=WF(\omega_2)\cup WF(\tilde{\omega}_2)$. Intersecting
with $\Gamma_2$ then gives the result.
\end{proof*}

The following result on the comparison of two generalised Hadamard
states is well known and lies at the basis of the renormalisation of
the stress-energy-momentum tensor in the free field case:
\begin{lemma}\label{lem_2compare}
For two generalised Hadamard states $\omega,\omega'$ we have that
$\omega_2-\omega_2'$ is smooth iff $\omega_{2-}-\omega_{2-}'$ is
smooth.
\end{lemma}
\begin{proof*}
We define $w(x_2,x_1):=(\omega_2-\omega_2')(x_2,x_1)$,
$\tilde{w}(x_2,x_1):=w(x_1,x_2)$ and $w_{2-}:=\frac{1}{2}(w-\tilde{w})$
and argue as in the proof of proposition \ref{prop2point}:
$WF(w)\subset\Gamma_2$, $WF(\tilde{w})\subset-\Gamma_2$ and hence
$WF(w)\cap WF(\tilde{w})=\emptyset$. It then follows from
$w-\tilde{w}=2w_{2-}$ that $WF(w_{2-})=WF(w)\cup WF(\tilde{w})$.
Now $WF(w_{2-})=\emptyset$ if and only if $WF(w)=\emptyset$, which
proves the statement.
\end{proof*}

\section{Generalised free fields in curved spacetime}\label{sec_GFF}

In this section we define a number of physical properties that the
state $\omega$ may satisfy and derive some easy results concerning
them:
\begin{definition}
The state $\omega$ is called \emph{causal} iff $\omega$ descends to a
state on $\Alg{U}_M/J$, where $J\subset\Alg{U}_M$ is the $^*$-ideal
generated by all elements of the form $f\otimes h-h\otimes f$ where
the supports of $f,h\in\Test_0(M)$ are causally disjoint.

A state $\omega$ satisfies the \emph{Klein-Gordon equation} with mass
$m$ and scalar curvature coupling $\xi$ iff $\omega$ descends to a
state on $\Alg{U}_M/J$, where $J\subset\Alg{U}_M$ is the $^*$-ideal
generated by all elements of the form $(\Box+m^2+\xi R)f$, where
$\Box$ is the d'Alembertian and $R$ the scalar curvature.

Given a bi-distribution $E$ on $M^{\times 2}$ we say that the state
$\omega$ is a \emph{generalised free field state with commutator} $E$
iff $\omega$ \emph{satisfies the commutation relations with commutator}
$E$, i.e.~iff $\omega$ descends to a state on $\Alg{U}_M/J$, where
$J\subset\Alg{U}_M$ is the $^*$-ideal generated by all elements of
the form $f\otimes h-h\otimes f-iE(f,h)I$.

A generalised free field state $\omega$ on a globally hyperbolic
spacetime is called a free field state iff it satisfies the
Klein-Gordon equation with mass $m$ and scalar curvature coupling
$\xi$ and $E=E_{m,\xi}$, the difference of the advanced and retarded
fundamental solutions of the Klein-Gordon equation
$(\Box+m^2+\xi R)\phi=0$.
\end{definition}
The first three properties above can be written equivalently in terms
of the represented field as:
\begin{eqnarray}
\left[\Phi^{\omega}(f),\Phi^{\omega}(h)\right]=0&&
\mathrm{supp}\ f\cap J(\mathrm{supp}\ h)=\emptyset,\label{eqn_causality}\\
(\Box+m^2+\xi R)\Phi^{\omega}=0,&&\label{eqn_KG}\\
\left[\Phi^{\omega}(f),\Phi^{\omega}(h)\right]=iE(f,h)I.&&\label{eqn_CCR}
\end{eqnarray}

We have chosen to allow very general distributions $E$ to appear in the
commutation relations in order to emphasise that their precise form does
not matter for our arguments. In particular, our commutation relations
need not imply causality and our arguments also hold for anti-commutation
relations. However, it is important that the commutator of two smeared
field operators is a scalar. Note that $E$ must be anti-symmetric,
$E(f,h)=-E(h,f)$ for all $f,h\in\Test_0(M)$, for the commutation relations
to make sense.

Instead of the free field commutator $E=E_{m,\xi}$ one can take for
example
$E(x_2,x_1):=\int_0^{\infty}\int E_{m,\xi}(x_2,x_1)f(m,\xi)d\xi dm$,
where $f$ is a compactly supported smooth function. In fact, in the
Wightman framework in Minkowski spacetimeone can use the K\"allen-Lehmann
representation of the two-point distribution to prove that $E$ must be of
this form for a suitable distribution $f(m)\delta(\xi)$. (We can take
$\xi=0$ because $R\equiv 0$ in Minkowski spacetime). Whether such a
result still holds in curved spacetime is not clear, because no suitable
replacement of the K\"allen-Lehmann is currently available. We hope to
return to these issues in more detail elsewhere \cite{Sanders2}.

It is worthwhile to note the following:
\begin{proposition}\label{prop_Had2point}
If $\omega$ is a generalised Hadamard state and a generalised free field
state with commutator $E$, then $E=-2i\omega_{2-}$,
$WF(E)\subset\Gamma_2\cup-\Gamma_2$ and
$WF(\omega_2)=WF(E)\cap\Gamma_2$.
\end{proposition}
\begin{proof*}
The first equality follows by applying $\omega$ to the commutation
relations (\ref{eqn_CCR}). The others follow from the last two items of
proposition \ref{prop2point}.
\end{proof*}

Note that $WF(E_{m,\xi})\subset\Gamma_2^{ll}\cup-\Gamma_2^{ll}$, so in this
case we must have $WF(\omega_2)\subset\Gamma_2^{ll}$.

\begin{corollary}
If a state $\omega$ satisfies the Klein-Gordon equation with parameters
$m,\xi$ on a globally hyperbolic spacetime $M$ and the commutation relations
with commutator $E_{m,\xi}$, then it is a generalised Hadamard state if and
only if it is a Hadamard state.
\end{corollary}
\begin{proof*}
The result of Radzikowski implies that a free field state on a globally
hyperbolic spacetime which satisfies the commutation relations with
commutator $E_{m,\xi}$ is a Hadamard state if and only if
$WF(\omega_2)=WF(E_{m,\xi})\cap\Gamma_2$. The result therefore follows
from proposition \ref{prop_Had2point}.
\end{proof*}

We see from proposition \ref{prop_Had2point} and lemma \ref{lem_2compare}
that for two generalised free field states $\omega$ and $\omega'$ with
commutator functions $E$ and $E'$ respectively, $\omega_2-\omega_2'$ is smooth
iff $E-E'$ is smooth. In general, however, even $E_{m,\xi}-E_{m',\xi'}$ will
not be smooth, even though both have the same wave front sets. Indeed, if
$E_{m,\xi}-E_{m',\xi'}$ is smooth and if we define $K_{x_2}:=\Box+m^2+\xi R$
acting on the variable $x_2$ and similarly for $K'$, then the following is
also smooth:
\begin{eqnarray}
&&K'_{x_2}(E_{m,\xi}-E_{m',\xi'})(x_2,x_1)=K'_{x_2}E_{m,\xi}(x_2,x_1)\nonumber\\
&=&(K'_{x_2}-K_{x_2})E_{m,\xi}(x_2,x_1)\nonumber\\
&=&((m')^2-m^2+\xi'R(x_2)-\xi R(x_2))E_{m,\xi}(x_2,x_1).\nonumber
\end{eqnarray}
Because $E_{m,\xi}(x_2,x_1)$ is singular whenever $x_1$ and $x_2$ can be
connected by a light-like geodesic, we would then have to have
$(m')^2-m^2+\xi'R(x_2)-\xi R(x_2)\equiv 0$. In general, however, this is not
the case.

We conclude this section by proving a useful equivalent characterisation of
generalised free fields in terms of the truncated $n$-point distributions:
\begin{proposition}\label{prop_wTsym}
A state $\omega$ is a generalised free field state iff all the truncated
$n$-point distributions $\omega^T_n$ with $n\not=2$ are symmetric in their
arguments.
\end{proposition}
In the case where we have anti-commutation relations instead of commutation
relations a similar proof shows that the truncated $n$-point distributions
are anti-symmetric for $n\not=2$.
\begin{proof*}
First assume that $\omega^T_n$ is symmetric for all $n\not=2$. For $n\ge 2$
we then use equation (\ref{eqn_deftrunc}) to see that for any
$1\le i<n$
\begin{eqnarray}
\omega_n(x_n,\ldots,x_1)-\omega_n(x_n,\ldots,x_i,x_{i+1},\ldots,x_1)&=&\nonumber\\
2\omega_{2-}(x_{i+1},x_i)\omega_{n-2}(x_n,\ldots,\hat{x}_{i+1},\hat{x}_i,
\ldots,x_1).&&\nonumber
\end{eqnarray}
Here we noted that most terms cancel out, either by the hypothesis or by the
fact that $i$ and $i+1$ are subsequent indices. The remaining terms have been
collected together using once again equation (\ref{eqn_deftrunc}). By definition
this equation means that $\omega$ satisfies the commutation relations with
commutator $E=-2i\omega_{2-}$.

For the opposite direction we assume that $\omega$ satisfies the commutation
relations (necessarily with $E=-2i\omega_{2-}$). We use similar arguments as
above to prove by induction that $\omega^T_n$ is symmetric for $n\ge 3$. (For
$n=0,1,2$ there is nothing to prove.)
\begin{eqnarray}
\omega^T_3(x_3,x_2,x_1)-\omega^T_3(x_2,x_3,x_1)&=&\nonumber\\
\omega_3(x_3,x_2,x_1)-\omega_3(x_2,x_3,x_1)-(\omega^T_2(x_3,x_2)
-\omega^T_2(x_2,x_3))\omega^T_1(x_1)&=&\nonumber\\
2\omega_{2-}(x_3,x_2)\omega_1(x_1)-2\omega_{2-}(x_3,x_2)\omega_1(x_1)&=&0.\nonumber
\end{eqnarray}
A similar result holds for the transposition in the indices $1$ and $2$, from which
the invariance under all permutations follows for $n=3$.
Next we consider $n>3$ and assume that the claim holds for all $\omega^T_{n'}$
with $0\le n'\le n-1$. Again it suffices to prove that $\omega^T_n(x_n,\ldots,x_1)$
is invariant under a transposition of the indices $i$ and $i+1$ for some
$1\le i\le n-1$, because such transpositions generate the group of all
permutations. Using the induction hypothesis we find similarly:
\begin{eqnarray}
\omega^T_n(x_n,\ldots,x_1)-\omega^T_n(x_n,\ldots,x_i,x_{i+1},\ldots,x_1)&=&
\nonumber\\
\omega_n(x_n,\ldots,x_1)-\omega_n(x_n,\ldots,x_i,x_{i+1},\ldots,x_1)&&\nonumber\\
-(\omega^T_2(x_{i+1},x_i)-\omega^T_2(x_i,x_{i+1}))\omega_{n-2}(x_n,\ldots,
\hat{x}_{i+1},\hat{x}_i,\ldots,x_1)&=&0.\nonumber
\end{eqnarray}
This completes the proof.
\end{proof*}
The previous proposition is reminiscent of, but certainly not equivalent to,
the result in \cite{Jost} that a vacuum state $\omega$ of a Wightman field theory
is causal if and only if the $n$-point distributions $\omega_n$, extended to
suitable complex domains, are symmetric in their arguments in those domains.
That result, however, uses the Bargmann-Hall-Wightman theorem, whereas our result
relies solely on elementary combinatorics (cf.~\cite{Jost} section 4.4,
\cite{Jost2,Hall+}).

Finally we note the following corollary of proposition
\ref{prop_wTsym}\footnote{We thank prof. Rehren for pointing this out to us.}:
\begin{corollary}\label{cor_qf}
A quasi-free state satisfies the commutation relations with commutator
$E=-2i\omega_{2-}$.
\end{corollary}
\begin{proof*}
By definition \ref{def_qf} of a quasi-free state $\omega^T_n$ is symmetric for
$n\not=2$.
\end{proof*}

\section{Equivalence of the Hamadard and microlocal spectrum conditions.}
\label{sec_npoint}

We now start our analysis of the singularities of higher $n$-point
distributions of a generalised free field state with a result that exploits
the positivity of the state.

\begin{proposition}\label{prop_positivity}
Let $\omega$ be a generalised Hadamard state and assume that for
$n\ge 1$ we have $(x_n,k_n;\ldots;x_1,k_1)\in WF(\omega_n)$. Then
$(x_1,k_1)\in V^{*+}\cup\mathcal{Z}$ and
$(x_n,k_n)\in V^{*-}\cup\mathcal{Z}$. In particular,
$WF(\omega_1)=\emptyset$.
\end{proposition}
\begin{proof*}
The positivity of $\omega$ implies
$\overline{\omega_n(f_n,\ldots,f_1)}=
\omega_n(\bar{f}_1,\ldots,\bar{f}_n)$, and hence the second
statement follows from the first. In fact, the positivity allows us to
perform the GNS-construction, which yields a representation
$\pi_{\omega}$ of $\Alg{U}_M$ on a Hilbert space
$\mathcal{H}_{\omega}$ by closable operators and a vector
$\Omega_{\omega}\in\mathcal{H}_{\omega}$ such that
$\omega(A)=\left<\Omega_{\omega},\pi_{\omega}(A)\Omega_{\omega}\right>$
for all $A\in\Alg{U}_M$. We can then define the
$\mathcal{H}_{\omega}$-valued distributions
$\phi_{m}(f_m,\ldots,f_1):=\pi_{\omega}(f_m\otimes\ldots\otimes f_1)
\Omega_{\omega}$ for all $m\in\N$. Using the inner product of
$\mathcal{H}_{\omega}$ we can write:
\begin{eqnarray}
\omega_n(f_n,\ldots,f_1)&=&\left<\phi_{n-1}(\bar{f}_2,\ldots,\bar{f}_n),
\phi_1(f_1)\right>,\nonumber\\
\omega_2(f_2,f_1)&=&\left<\phi_1(\bar{f}_2),\phi_1(f_1)\right>.\nonumber
\end{eqnarray}
The calculus of Hilbert space-valued distributions (see
e.g.~\cite{Strohmaier+} proposition 2.2 or \cite{Sanders} theorem
A.1.6) now means that $(x_n,k_n;\ldots;x_1,k_1)\in WF(\omega_n)$
implies $(x_1,k_1)\in WF(\phi_1)\cup\mathcal{Z}$ and if $k_1\not=0$
then $(x_1,-k_1;x_1,k_1)\in WF(\omega_2)$. The conclusion follows
from the assumption that $WF(\omega_2)\subset \Gamma_2$.
\end{proof*}

Proposition \ref{prop_positivity} has some nice consequences in the
case of generalised free fields:
\begin{theorem}\label{thm_Tsmooth}
Let $\omega$ be a generalised Hadamard state which is also a
generalised free field state. Then $\omega^T_2-\omega_2$ and
$\omega^T_n$ for all $n\not=2$ are smooth functions.
\end{theorem}
\begin{proof*}
From proposition \ref{prop_positivity} and equation
(\ref{eqn_deftrunc}) we see that
$(x_n,k_n;\ldots;x_1,k_1)\in WF(\omega^T_n)$ implies
$(x_1,k_1)\in V^{*+}\cup\mathcal{Z}$ and
$(x_n,k_n)\in V^{*-}\cup\mathcal{Z}$. However, because $\omega$ is
a generalised free field state all truncated $n$-point distributions
with $n\not=2$ are symmetric by proposition \ref{prop_wTsym}.
This means that each $(x_i,k_i)$ must be in
$(V^{*+}\cup\mathcal{Z})\cap(V^{*-}\cup\mathcal{Z})=\mathcal{Z}$, i.e.
$k_i=0$. It follows that $WF(\omega^T_n)=\emptyset$ and hence
$\omega^T_n$ is smooth for $n\not=2$. The result for $n=2$ follows
from $\omega_2-\omega^T_2=\omega_1\otimes\omega_1$.
\end{proof*}

\begin{corollary}\label{cor_mSC}
Let $\omega$ be a generalised Hadamard state which is also a generalised
free field state. Then $\omega$ satisfies the microlocal spectrum
condition with smooth, resp.~causal, resp.~light-like immersions if
$WF(\omega_{2-})\subset\Gamma_2^{\bullet}$, where $^{\bullet}$ denotes no
superscript, resp.~$c$, resp.~$ll$. More precisely, for each point in
$WF(\omega_n)$ we can find an instantiating graph $G\in\mathcal{G}_n$
which is a disconnected union of graphs in $\mathcal{G}_2$ that
instantiate points in $WF(\omega_2)=WF(E)\cap\Gamma_2$.
\end{corollary}
\begin{proof*}
This follows immediately from theorem \ref{thm_Tsmooth}, equation
(\ref{eqn_deftrunc}) and the properties of the cones $\Gamma_n^{\bullet}$
in proposition \ref{propGamma}.
\end{proof*}

The singularity structure that we derived in theorem \ref{thm_Tsmooth}
and corollary \ref{cor_mSC} is what one would expect of quasi-free states,
because of equation (\ref{eqn_deftrunc}) (see \cite{Brunetti+1}). It is
nice to see that this form persists when the state is only required to
satisfy scalar commutation relations. Analogous results also hold in the
analytic case, for vector-valued fields and in the case of anti-commutation
relations.

\cite{Brunetti+1} describes a point in $T^*M^5\setminus\mathcal{Z}$ that is
not in $\Gamma_5^c$ and wonders whether such a point can be in the wave
front set of the 5-point distribution of a state. We have just proved that
for generalised free fields this possibility is excluded. Moreover,
our result also implies that the $\mu$SC with light-like curves includes
more than just free fields and their Wick powers \cite{Brunetti+1},
namely generalised free fields with any suitable commutator function. (For
the existence of a sufficiently large class of such fields in curved
spacetime we refer to \cite{Sanders2}.)

An easy consequence of the analytic case of theorem \ref{thm_Tsmooth} is
the following characterisation of generalised free field states:
\begin{proposition}
Let $\omega$ be a causal state satisfying the A$\mu$SC. Then $\omega$ is
a generalised free field state if and only if $\omega^T_n$ is analytic for
all $n\not=2$.
\end{proposition}
\begin{proof*}
If $\omega$ is a generalised free field state the conclusion follows from
the analytic version of theorem \ref{thm_Tsmooth}. For the converse we use
causality to prove by induction on $n$ that every $\omega^T_n$ is symmetric
when all arguments are space-like separated. Analytic continuation for
$n\not=2$ then proves their symmetry everywhere and we may then apply
proposition \ref{prop_wTsym}.
\end{proof*}

As another easy result we show that the class of generalised Hadamard
states of a generalised free field is closed under operations:
\begin{proposition}\label{prop_closed}
Let $\omega$ be a generalised Hadamard and generalised free field state on
$\Alg{U}_M$ and let $A\in\Alg{U}_M$ be any operator such that
$\omega(A^*A)=1$. Then the state $\omega^A$, defined by
$\omega^A(B):=\omega(A^*BA)$, is a generalised Hadamard and generalised
free field state on $\Alg{U}_M$.
\end{proposition}
Notice that for given $A$ the expression $\omega(A^*BA)$ may involve
arbitrary high $n$-point distributions, depending on the choice of
$B$, so without estimate on the wave front sets of higher $n$-point
distributions this result sounds rather surprising.
\begin{proof*}
We may write $A=\sum_{i=1}^nf^{(i)}_i\otimes\ldots\otimes f^{(i)}_1$
for some $n$ and $f^{(i)}_j\in\Test_0(M)$. The two-point distribution of
$\omega^A$ is then a sum of terms of the form
\[
\omega_{i+k+2}\left(\overline{f^{(i)}_1},\ldots,\overline{f^{(i)}_i},
x_2,x_1,f^{(k)}_k,\ldots,f^{(k)}_1\right)
\]
which are distributions in $x_1,x_2$. The wave front set of each such
term can be estimated using standard arguments (see \cite{Hoermander}
theorem 8.2.12) as a subset of
\[
\left\{(x_2, k_2;x_1,k_1)|\ 
(y_1,0;\ldots;y_i,0;x_2,k_2;x_1,k_1;z_k,0;\ldots;z_1,0)\in 
WF(\omega_{i+k+2})\right\} 
\]
which is a subset of $\Gamma_2$. The wave front set of a sum of such
terms is also contained in $\Gamma_2$ and therefore $\omega^A$ is a
generalised Hadamard state. That it is a generalised free field state
follows from equation (\ref{eqn_CCR}).
\end{proof*}

To close this section we prove the following lemma on the comparisons
of the $n$-point distributions of two states, generalising lemma
\ref{lem_2compare}.
\begin{lemma}
Consider two generalised Hadamard states $\omega,\omega'$, which both
satisfy commutation relations with the same commutator $E$ such that
$WF(E)\not=\emptyset$. For any $n\ge 0$ we have that
$\omega_{n+2}-\omega'_{n+2}$ is smooth if and only if
$\omega_n\equiv \omega_n'$.
\end{lemma}
\begin{proof*}
The case $n=0$ follows from lemma \ref{lem_2compare}. For $n\ge 1$
we first suppose that $\omega_n\equiv\omega_n'$. For any index
$1\le i<n$ we then have
$(\omega_{n+2}-\omega_{n+2}')(x_{n+2},\ldots,x_1)=
(\omega_{n+2}-\omega_{n+2}')(x_{n+2}
,\ldots,x_i,x_{i+1},\ldots,x_1)$,
where we swapped the indices $i$ and $i+1$ and the commutator terms
vanish by the assumption. We can therefore permute indices ad
lib.~and in this way we derive
$(\omega_{n+2}-\omega_{n+2}')(x_{n+2},\ldots,x_1)=
(\omega_{n+2}-\omega_{n+2}')(x_1,\ldots,x_{n+2})$. Using the
assumption that both states are generalised Hadamard states and items
two and four of proposition \ref{propGamma} we find that
$WF(\omega_{n+2}-\omega_{n+2}')\subset \Gamma_{n+2}\cap-\Gamma_{n+2}=
\emptyset$. This proves that $\omega_{n+2}-\omega_{n+2}'$ is smooth.

For the opposite direction we assume that $\omega_{n+2}-\omega_{n+2}'$
is smooth and we let the symbol $\sim$ denote equality modulo terms
$w$ such that
$WF(w)\cap T^*M\times V^{*+}\times T^*M\times\ldots\times T^*M
=\emptyset$, i.e.~we are interested in the direction of the covectors
in the $n+1$st slot (from the right). Using the expressions for
$\omega_{n+2}$ and $\omega_{n+2}'$ in terms of truncated $n$-point
distributions (\ref{eqn_deftrunc}) we compute:
\begin{eqnarray}
0&\sim&\omega_{n+2}-\omega_{n+2}'\sim \omega_2\otimes\omega_n-
\omega_2'\otimes\omega_n'\nonumber\\
&\sim&\omega_2\otimes\omega_n-\omega_2\otimes\omega_n'
=\omega_2\otimes(\omega_n-\omega_n'),\nonumber
\end{eqnarray}
where we used the result for $n=0$ to get to the last line.
If there is a point $(x_n,\ldots,x_1)$ where
$w_n:=\omega_n-\omega_n'\not= 0$ then we can find
test-functions $f_1,\ldots,f_n$ such that $c:=w_n(f_n,\ldots,f_1)\not=0$,
which leads to a contradiction as follows. Notice that
\begin{eqnarray}
WF(\omega_2)&=&WF(c\cdot\omega_2)=WF(\omega_2\cdot w_n(f_n,\ldots,f_1))
\nonumber\\
&\subset&\left\{(x_{n+2},k_{n+2};x_{n+1},k_{n+1})|\
\mathrm{\ for\ some\ } x_i\in\mathrm{supp}(f_i),\ i=1,\ldots,n
\right.\nonumber\\
&&\left.(x_{n+2},k_{n+2};x_{n+1},k_{n+1};x_n,0;\ldots;x_1,0)\in
WF(\omega_2\otimes w_n)\right\},\nonumber
\end{eqnarray}
by theorem 8.2.12 of \cite{Hoermander}. Because
$\omega_2\otimes w_n\sim 0$ and because $\omega_2$ is a generalised
Hadamard state we find that $WF(\omega_2)=\emptyset$. However, by
proposition \ref{prop_Had2point} this implies that
$WF(E)\cap\Gamma_2=\emptyset$ and hence $WF(E)\cap-\Gamma_2=\emptyset$
and $WF(E)=\emptyset$. This contradicts the assumption on $E$, so we
must have $w_n\equiv 0$.
\end{proof*}
The same statement still holds when the commutators $E$ and $E'$ of
the two states differ by a smooth function.

\section{Two theorems generalised to curved spacetimes}\label{sec_2thm}

We now discuss the generalisation of two theorems from Wightman field
theory to curved spacetimes, illustrating the strength and the limitations
of the commutation relations in that setting. First we generalise a result
due to Borchers and Zimmermann \cite{Borchers+} concerning the
self-adjointness of field operators. Then we consider the generalisation
of (a weak form of) the Jost-Schroer theorem.

The result of \cite{Borchers+} gives a sufficient condition for the symmetric
operator $\Phi^{\omega}(f)$ with a given $f\in\Test_0(M,\R)$ to be
self-adjoint. To discuss its generalisation we recall the following
notion:
\begin{definition}
A vector $\psi$ in a Hilbert space $\mathcal{H}$ is an
\emph{analytic vector} for a (possibly unbounded) linear operator
$T$ on $\mathcal{H}$ iff the series
$\sum_{n=0}^{\infty}\frac{\|T^n\psi\|}{n!}z^n$ has a non-zero
radius of convergence. (In particular we require that $\psi$ is
in the domain of each $T^n$.)
\end{definition}
Notice that for a bounded linear operator $T$ all vectors are
analytic. The following elementary lemma is adapted from \cite{Borchers+}:
\begin{lemma}\label{lem_charanalytic}
For a vector $\psi$ in the Hilbert space $\mathcal{H}$ and a
symmetric linear operator $T$ on $\mathcal{H}$ the following are
equivalent:
\begin{itemize}
\item $\psi$ is analytic for $T$,
\item there is a constant $c>0$ such that $\|T^n\psi\|\le n!c^n$,
\item $\sum_{n=0}^{\infty}\frac{|\langle\psi,T^n\psi\rangle|}{n!}z^n$
has a non-zero radius of convergence,
\item there is a constant $c>0$ such that
$|\langle\psi,T^n\psi\rangle|\le n!c^n$.
\end{itemize}
\end{lemma}
\begin{proof*}
If $\psi$ is analytic for $T$ then the second condition follows by
choosing $z>0$ suitably small. Similarly the third condition implies the
fourth. The first condition also implies the third, for if $\psi$ is
analytic for $T$ then
\[
\sum_{n=0}^{\infty}\frac{|\langle\psi,T^n\psi\rangle|}{n!}|z|^n\le
\|\psi\|\sum_{n=0}^{\infty}\frac{\|T^n\psi\|}{n!}|z|^n
\]
so the left-hand side has a finite radius of convergence. Similarly, the
second condition implies the fourth. Finally we show that the fourth
condition implies the first. For this we note
that $\|T^n\psi\|^2=\langle\psi,T^{2n}\psi\rangle\le c^{2n}(2n)!$
because $T$ is symmetric. From $(2n)!\le(2^nn!)^2$ we deduce that
$\frac{\|T^n\psi\|}{n!}\le(2c)^n$ and hence that $\psi$ is analytic.
\end{proof*}

For a Wightman field theory in Minkowski spacetime Borchers and
Zimmermann \cite{Borchers+} used causality and the the Reeh-Schlieder
theorem to prove that a field operator $\pi_{\omega}(\Phi(f))$ is
self-adjoint as soon as the vacuum vector $\Omega_{\omega}$ is analytic.
An analogous proof can be given in curved spacetime, whenever the state
$\omega$ is causal and has the Reeh-Schlieder property, i.e.~the GNS-vector
$\Omega_{\omega}$ is cyclic for all local algebras. The latter can be
ensured e.g.~by imposing the A$\mu$SC (see
\cite{Strohmaier+,Sanders1}), but unfortunately it is not clear whether
all analytic spacetimes admit states satisfying the A$\mu$SC, or whether
all (smooth) spacetimes have states with the Reeh-Schlieder property.
We now prove that the conclusion of Borchers and Zimmermann can also be
obtained without recourse to the Reeh-Schlieder theorem if we assume that
the state is a generalised free field state. For this purpose we adapt
an idea of Nelson \cite{Nelson}.
\begin{theorem}\label{thm_BZ}
If $\omega$ is a generalised free field state on $\Alg{U}_M$ with some
commutator $E$ and $\Omega_{\omega}$ is an analytic vector for
$\Phi^{\omega}(f)$ for some $f\in\Test_0(M,\R)$, then all vectors
$\pi_{\omega}(A)\Omega_{\omega}$ with $A\in\Alg{U}_M$ are analytic vectors
for $\Phi^{\omega}(f)$ and this operator is essentially self-adjoint.
\end{theorem}
\begin{proof*}
First assume that $\psi\in\pi_{\omega}(\Alg{U}_M)\Omega_{\omega}$ is an
analytic vector for $\Phi^{\omega}(f)$ for given $f\in\Test_0(M,\R)$. For
any $h\in\Test_0(M)$ we will prove that $\Phi^{\omega}(h)\psi$ is an
analytic vector for $\Phi^{\omega}(f)$. To see this we note that for
$n\ge 1$ we have
\[
\Phi(f)^n\Phi(h)=\Phi(h)\Phi(f)^n+niE(f,h)\Phi(f)^{n-1},
\]
which may easily be proved by induction. Using this we compute:
\begin{eqnarray}
\|\Phi^{\omega}(f)^n\Phi^{\omega}(h)\psi\|^2&=&
\langle\Phi^{\omega}(h)\psi,\Phi^{\omega}(f)^{2n}\Phi^{\omega}(h)\psi
\rangle\nonumber\\
&=&\langle\Phi^{\omega}(\overline{h})\Phi^{\omega}(h)\psi,
\Phi^{\omega}(f)^{2n}\psi\rangle\nonumber\\
&&+2niE(f,h)\langle\Phi^{\omega}(h)\psi,\Phi^{\omega}(f)^{2n-1}\psi
\rangle\nonumber\\
&\le&\|\Phi^{\omega}(\overline{h})\Phi^{\omega}(h)\psi\|\cdot
\|\Phi^{\omega}(f)^{2n}\psi\|\nonumber\\
&&+2n|E(f,h)|\cdot\|\Phi^{\omega}(h)\psi\|\cdot
\|\Phi^{\omega}(f)^{2n-1}\psi\|\nonumber\\
&\le& c\|\Phi^{\omega}(f)^{2n}\psi\|+2nc\|\Phi^{\omega}(f)^{2n-1}\psi\|,
\nonumber
\end{eqnarray}
where the constant $c>0$ may depend on $f$ and $h$, but not on
$n$. The assumption that $\psi$ is analytic then implies that
(see lemma \ref{lem_charanalytic})
\[
\|\Phi^{\omega}(f)^n\Phi^{\omega}(h)\psi\|^2\le c(c')^{2n}(2n)!+
2nc(c')^{2n-1}(2n-1)!\le C^{2n}(2n)!\nonumber
\]
for suitable constants $c',C>0$. Using the estimate $(2n)!\le (2^nn!)^2$
we then find that $\|\Phi^{\omega}(f)^n\Phi^{\omega}(h)\psi\|\le (2C)^nn!$,
which implies by lemma \ref{lem_charanalytic} that
$\Phi^{\omega}(h)\psi$ is an analytic vector for $\Phi^{\omega}(f)$.

Now assume that $\Omega_{\omega}$ is an analytic vector for
$\Phi^{\omega}(f)$. We can then repeatedly apply the result of the previous
paragraph to prove that any vector of the form 
$\Phi^{\omega}(h_m)\cdots\Phi^{\omega}(h_1)\Omega_{\omega}$ is an analytic
vector. Because the set of analytic vectors for a given operator is a linear
space every vector in $\pi_{\omega}(\Alg{U}_M)\Omega_{\omega}$ is analytic.
This provides a dense set of analytic vectors, so we can apply Nelson's
theorem (\cite{Nelson} lemma 5.1) to conclude that $\Phi^{\omega}(f)$ is
essentially self-adjoint.
\end{proof*}

The analyticity of $\Omega_{\omega}$ can be formulated conveniently in terms
of the $n$-point distributions by lemma \ref{lem_charanalytic} and in terms
of the truncated $n$-point distributions too (for a proof we refer to
\cite{Borchers+}):
\begin{proposition}\label{prop_AwT}
$\Omega_{\omega}$ is an analytic vector for $\Phi^{\omega}(f)$ if and only
if there is a $d>0$ such that $|\omega_n^T(f^{\otimes n})|<n!d^n$ for
all $n\in\N$.
\end{proposition}

The condition of the previous theorem may not always be satisfied, as we will
now illustrate with the following
\begin{example}
In Minkowski spacetime we will construct a translation invariant free field
state $\tilde{\omega}$ which satifies the A$\mu$SC, but whose $GNS$-vector
$\Omega_{\tilde{\omega}}$ is not analytic for any non-zero smeared field
operator $\Phi^{\tilde{\omega}}(f)$, $f\in\Test_0(M_0,\R)$. (We will not
discuss the question whether these operators are essentially self-adjoint.)

Let $\omega$ denote the Minkowski vacuum state with two-point distribution $\omega_2$.
We set $w_2(x_2,x_1):=\int e^{-ik\cdot(x_1-x_2)} e^{-k_0^2}\delta(k^2-m^2)dk$, which
is an analytic, real-valued, symmetric and translation invariant bi-solution of the
Klein-Gordon equation of positive type. Next we define the two-point distributions
$\omega^j_2:=e^jw_2+\omega_2$ for each $j\in\N$ and we note that the anti-symmetric
part is $\omega^j_{2-}=\omega_{2-}$. Each $\omega^j_2$ defines a quasi-free state
$\omega^j$ on the Weyl-algebra (see \cite{Kaersten}) and hence also on the
Borchers-Uhlmann algebra, because a quasi-free state is regular
(cf.~proposition \ref{prop_AwT}). Each of the states $\omega^j$ is a translation
invariant, Hadamard, free field state satifying the A$\mu$SC. (Note however that
they are not Lorentz-invariant, because $w_2$ is not Lorentz invariant.)

Now we define the state $\tilde{\omega}$ by
$\tilde{\omega}:=e^{-1}\sum_{j=0}^{\infty}\frac{1}{j!}\omega^j$. Note that
$\tilde{\omega}(A^*A)\ge 0$ and $\tilde{\omega}_0(I)=1$, so it is indeed a state. It
follows from the properties of the $\omega^j$ that $\tilde{\omega}$ is
translation invariant and that it is a free field state. To see that $\tilde{\omega}$
is continuous we note that $\tilde{\omega}_{2n-1}=0$ for $n\in\N$ and that for all
$n,N\in\N$:
\[
e^{-1}\sum_{j=0}^N\frac{1}{j!}\omega^j_{2n}=e^{-1}\sum_{P\in\mathcal{P}_n}\sum_{j=0}^N
\frac{1}{j!}(e^jw_2+\omega_2)^{\otimes n}\circ\pi_P
\]
\[
=e^{-1}\sum_{P\in\mathcal{P}_n}\sum_{k=0}^n\sum_{j=0}^N
\frac{e^{kj}}{j!}\left(w_2^{\otimes k}\otimes\omega_2^{\otimes (n-k)}+\ldots+
\omega_2^{\otimes (n-k)}\otimes w_2^{\otimes (n-k)}\right)\circ\pi_P,
\]
where the operation $\pi_P$ denotes the permutation that corresponds to the
partition $P$ of the set $\left\{1,\ldots,n\right\}$ (see equation (\ref{eqn_deftrunc})
and definition \ref{def_qf}) and the dots in the last line indicate all the different
orderings of the factors $w_2$ and $\omega_2$. Taking the limit we see that the sum over
$j$ converges so that
\begin{equation}\label{eqn_w2distr}
\tilde{\omega}_{2n}=\sum_{P\in\mathcal{P}_n}\sum_{k=0}^n e^{e^k-1}
\left(w_2^{\otimes k}\otimes\omega_2^{\otimes (n-k)}+\ldots+
\omega_2^{\otimes (n-k)}\otimes w_2^{\otimes (n-k)}\right)\circ\pi_P,
\end{equation}
which exhibits $\tilde{\omega}_{2n}$ as a finite sum of distributions. It also follows
from equation (\ref{eqn_w2distr}) that $\tilde{\omega}$ satisfies the A$\mu$SC.

Finally we prove that $\Omega_{\tilde{\omega}}$ is not an analytic vector for any
non-zero $\Phi^{\tilde{\omega}}(f)$ with $f\in\Test_0(M_0,\R)$. For suppose that
$\Omega_{\tilde{\omega}}$ is an analytic vector for a given
$\Phi^{\tilde{\omega}}(f)$. By lemma \ref{lem_charanalytic} there is a constant
$c>0$ such that
\[
c^{2n}(2n)!\ge\tilde{\omega}_{2n}(f^{2n})\ge\frac{(2n)!}{2^nn!}e^{e^n-1}w_2(f,f)^n,
\]
where we used equation (\ref{eqn_w2distr}) and the positive type of $\omega_2$ and
$w_2$ for the last inequality. Using  $\ln n\le n$ we find $n!\le n^n\le e^{n^2}$
and hence
\[
c^{2n}\ge \left(\frac{w_2(f,f)}{2}\right)^n e^{n^3/6-n^2}.
\]
If $w_2(f,f)\not=0$ we can take logarithms on both sides and let $n\rightarrow\infty$
to find a contradiction. If $w_2(f,f)=0$, on the other hand, we use the positivity
and the support of $w_2$ to deduce that $\omega_2(f,f)=0$ too and hence
$\tilde{\omega}_2(f,f)=0$. This means that $\Phi^{\tilde{\omega}}(f)$ annihilates
$\Omega_{\tilde{\omega}}$ and it commutes with all other smeared field operators, so
that $\Phi^{\tilde{\omega}}(f)=0$ (cf.~the proof of proposition \ref{prop_JS1} below).
\end{example}

Now we turn to an analogue of the Jost-Schroer theorem (see \cite{Jost2},
\cite{Pohlmeyer},\cite{Federbush+}), which provides a way to recognise free
field states. In the Wightman framework this theorem says that any state whose
two-point distribution is that of a free field must be a free field
state\footnote{A related result, due to Greenberg \cite{Greenberg}, says
that a state must be a generalised free field state if the K\"allen-Lehmann
representation of the two-point distribution
\[
\omega_{2-}(x_2,x_1)=\int \rho_{KL}(m^2)\omega^m_{2-}(x_2,x_1)dm^2
\]
in terms of the free field commutator functions of mass $m$, $\omega^m_{2-}$,
has a positive measure $\rho_{KL}$ whose support satisfies certain
restrictions. In the Wightman framework every $\omega_{2-}$ allows a
K\"allen-Lehmann representation, but in curved spacetime such a tool is not
available, so at present it makes no sense to consider the generalisation of
this result. Moreover, our current strategy of weakening the Wightman axioms
and assuming commutation relations instead would render the statement
trivial.}. (Recall that this means it satisfies the Klein-Gordon equation and
the canonical commutation relations.)

As before we can prove our result by using commutation relations to replace
the analyticity that is due to the spectrum condition of the Wightman axioms.
Note, however, that this makes part of the result, namely the proof of the
commutation relations, trivial. The following is therefore a generalisation
of a very weak form of the Jost-Schroer theorem:
\begin{proposition}\label{prop_JS1}
Let $\omega$ be a generalised free field state and assume that $\omega_2$ is
the two-point distribution of a free-field state, i.e.~it satisfies the
Klein-Gordon equation for some mass $m$ and scalar curvature coupling $\xi$
and $\omega_{2-}=\frac{i}{2}E_{m,\xi}$. Then $\omega$ is a free field state.
\end{proposition}
(The same result also works for other linear partial differential operators.)
\begin{proof*}
Let $K$ denote the Klein-Gordon operator with mass $m$ and coupling $\xi$.
For any $f\in\Test_0(M)$ we have $KG\Phi(f)=\Phi(KGf)$, because
the Klein-Gordon operator is formally self-adjoint. This implies that
\[
|\omega_n(f_n,\ldots,f_2,Kf_1)|\le\|\Phi^{\omega}(\overline{f}_2)\cdots
\Phi^{\omega}(\overline{f}_n)\Omega_{\omega}\|\cdot
\|\Phi^{\omega}(Kf_1)\Omega_{\omega}\|=0,\nonumber
\]
because $\|\Phi^{\omega}(Kf_1)\Omega_{\omega}\|^2=\omega_2(K\bar{f}_1,Kf_1)=0$.
Therefore every $\omega_n$ satisfies the Klein-Gordon equation in the first
(rightmost) argument. One proves by induction that the same is then true for
$\omega^T_n$, using equation (\ref{eqn_deftrunc}). For a generalised free field
state we can then apply proposition \ref{prop_wTsym} and find that $\omega^T_n$
satisfies the Klein-Gordon equation in all arguments for $n\not=2$. For $n=2$
this is true by the assumption on $\omega_2$. Using equation (\ref{eqn_deftrunc})
once more shows that the $\omega_n$ satisfy the Klein-Gordon equation in all
arguments, which completes the proof.
\end{proof*}

Alternatively we could drop the assumption that $\omega$ is a generalised free
field state and require causality and the A$\mu$SC (or the Reeh-Schlieder
property) instead. This certainly allows us to prove that $\omega$ satisfies
the Klein-Gordon equation as follows:
\begin{proposition}\label{prop_JS2}
Let $\omega$ be a causal state satisfying the A$\mu$SC. If $KG_x\omega_2(x,y)=0$
then $\omega$ satisfies the Klein-Gordon equation.
\end{proposition}
\begin{proof*}
By A$\mu$SC $\omega$ has the Reeh-Schlieder property, i.e.~$\Omega_{\omega}$ is
a cyclic vector for every local algebra \cite{Strohmaier+}. Now
$\Phi^{\omega}(KGf)$ annihilates $\Omega_{\omega}$ for every $f\in\Test_0(M)$
and $\Phi^{\omega}(KGf)\cdot\pi_{\omega}(B)\Omega_{\omega}=0$ for any $B$ that
commutes with $\Phi(f)$. By causality and the Reeh-Schlieder property we
conclude that $\Phi^{\omega}(KGf)$ annihilates a dense set of vectors and hence
$\Phi^{\omega}(KGf)=0$ (because the operator is closable).
\end{proof*}
Note, however, that it is not at all clear whether the state also satisfies
the canonical commutation relations. The proof of \cite{Pohlmeyer}, e.g., uses
Poincar\'e invariance, the full strength of the spectrum condition and the
uniqueness of the vacuum\footnote{In this connection it should also be noted that
generalised free fields need not have the time-slice property, so then the
commutation relations cannot be proved in curved spacetime via a
spacetime-deformation argument as in \cite{Verch}.}. We will not investigate
what other assumptions are necessary to recover the strong version of the
Jost-Schroer theorem, but for completeness we do provide the following:
\begin{example}
We construct a state satisfying the assumptions of proposition \ref{prop_JS2}
with the canonical commutator function, but which is not a generalised free
field state. For this purpose we let $\omega^1$ denote the quasi-free state on
Minkowski spacetime with $\omega^1_2=2\omega^0_2$, where $\omega^0$ is the
Minkowski vacuum. We let $\omega^2$ be the state with $\omega^2_n=0$ for all
$n>0$ and we note that the mixed state $\omega^3:=\frac{1}{2}(\omega^1+\omega^2)$
serves our purpose by considering the four-point distribution:
\[
\omega^3_4(x_4,x_3,x_2,x_1)-\omega^3_4(x_3,x_4,x_2,x_1)=
2i\omega^3_{2-}(x_4,x_3)\omega^3_2(x_2,x_1).
\]
\end{example}

\section{Applications and outlook}\label{sec_appl}

\cite{Kay} already mentions the class of Hadamard states whose truncated
$n$-point distributions are smooth functions for all $n\not=2$ as an
interesting class. Later \cite{Hollands+1,Hollands+2} discuss perturbation
theory by constructing an extended $^*$-algebra of Wick powers and
time-ordered products of a free field and find that the continuous states
on this algebra are exactly the Hadamard states of this class. Our theorem
\ref{thm_Tsmooth} shows that the condition on the truncated $n$-point
distributions is automatically satisfied for (generalised) free fields due
to the scalar commutation relations, so the class of \emph{all} Hadamard
states is the class of interest for perturbative quantum field theory.
Furthermore, corollary \ref{cor_mSC} shows that for a generalised free
field any generalised Hadamard state satisfies the microlocal spectrum
condition and proposition \ref{prop_closed} tells us that the class of
generalised Hadamard states is closed under operations, which is useful to
know from a fundamental point of view. Our theorem \ref{thm_Tsmooth} and
corollary \ref{cor_mSC} could find further applications in perturbative
quantum field theory around a generalised free field, rather than around a
free field. Such an approach has been suggested in \cite{Duetsch+} as a way
to gain insight in the AdS-CFT correspondence.

Concerning the strength of the assumption that a state is a generalised
free field state we have discussed the generalisation of two results from
the Wightman framework to curved spacetimes. We showed that in some
circumstances our assumption can replace the existing arguments based on
analyticity, as in theorem \ref{thm_BZ} that generalised a result of
Borchers and Zimmerman. For the Jost-Schroer theorem the situation was
more delicate: a weak form of this theorem can be proved in curved
spacetimes by assuming that a state is a generalised free field state.
However, it is not known if one can prove that a state is a
(generalised) free field under suitable circumstances without assuming
commutation relations in the first place.

Finally we note that the proofs we used were all elementary applications
of the calculus of wave front sets of (Hilbert space-valued) distributions
and the combinatorics of (truncated) $n$-point distributions. Both can
be generalised to vector-valued fields and to anti-commutation
relations in a straightforward manner (see e.g.~\cite{Sanders} proposition
4.2.17 for the result that a Hadamard state of the free Dirac field
satisfies the microlocal spectrum condition).

${}$\\[15pt]
{\bf Acknowledgements}\\[6pt]
I would like to thank Chris Fewster, Bernard Kay, Karl-Henning Rehren and
Pedro Lauridsen Ribeiro for helpful suggestions and discussions. This
research was supported by the German Research Foundation (Deutsche
Forschungsgemeinschaft (DFG)) through the Institutional Strategy of the
University of G\"ottingen and the Graduiertenkolleg 1493 ''Mathematische
Strukturen in der modernen Quantenphysik''. The results of section
\ref{sec_npoint} were obtained during the preparation of my PhD thesis at
the University of York.


\thebibliography{              }

\bibitem{Borchers+}
H.-J. Borchers and W. Zimmermann, \emph{On the self-adjointness of field
operators}, Nuovo Cimento (10) \textbf{31} (1964), 1047--1059

\bibitem{Brunetti+1}
R. Brunetti, K. Fredenhagen and M. K\"ohler, \emph{The microlocal
spectrum condition and Wick polynomials of free fields on curved
spacetimes}, Commun. Math. Phys. \textbf{180} (1996), 633--652

\bibitem{Duetsch+}
M. D\"utsch and K.-H. Rehren, \emph{Generalized free fields and the
AdS-CFT correspondence}, Ann. Henri Poincaré \textbf{4} (2003), 613--635

\bibitem{Federbush+}
P.G. Federbush and K.A. Johnson, \emph{Uniqueness property of the two-fold
vacuum expectation value}, Phys. Rev. \textbf{120} (1960), 1926

\bibitem{Greenberg}
O.W. Greenberg, \emph{Heisenberg fields which vanish on domains of momentum space},
J. Mathematical Phys. \textbf{3} (1962), 859--866 
 
\bibitem{Hall+}
D. Hall and A.S. Wightman, \emph{A theorem on invariant analytic
functions with applications to relativistic quantum field theory},
Mat.-Fys. Medd. Danske Vid. Selsk. \textbf{31} (1957), 1--41 

\bibitem{Hollands+1}
S. Hollands and R.M. Wald, \emph{Local Wick polynomials and time
ordered products of quantum fields in curved spacetime},
Commun. Math. Phys. \textbf{223} (2001), 289--326

\bibitem{Hollands+2}
S. Hollands and W. Ruan, \emph{The State Space of Perturbative Quantum
Field Theory in Curved Spacetimes},
Ann. Henri Poincar\'e \textbf{3} (2002), 635--657

\bibitem{Hoermander}
L. H\"ormander, \emph{The analysis of linear partial differential
operators I}, Springer, Berlin (2003)

\bibitem{Jost}
R. Jost, \emph{The general theory of quantized fields},
American Mathematical Soc., Providence, RI (1965)

\bibitem{Jost2}
R. Jost, \emph{Properties of Wightman functions},
in: \emph{Lectures on field theory and the many-body problem},
pp. 127--145, Academic Press, New York (1961)

\bibitem{Kaersten}
F. K\"arsten, \emph{Klassifikation der unit\"ar invarianten regul\"aren Zust\"ande
der Weylalgebra der CCR \"uber einem separablen Hilbertraum},
Report Math \textbf{06} (1989)
 
\bibitem{Kay}
B.S. Kay, \emph{Quantum field theory in curved spacetimes},
in \emph{Mathematical physics X} (Proceedings, Leipzig, Germany 1991),
K. Schm\"udgen (ed.), Springer, Berlin (1992)

\bibitem{Nelson}
E. Nelson, \emph{Analytic Vectors},
Ann. of Math. \textbf{70} (1959), 572--615

\bibitem{Pohlmeyer}
K. Pohlmeyer, \emph{The Jost-Schroer theorem for zero-mass fields},
Commun. Math. Phys. \textbf{12} (1969), 204--211

\bibitem{Radzikowski}
M.J. Radzikowski, \emph{Micro-Local Approach to the Hadamard Condition
in Quantum Field Theory on Curved Space-Time},
Commun. Math. Phys. \textbf{179} (1996), 529--553

\bibitem{Sanders1}
K. Sanders, \emph{On the Reeh-Schlieder Property in Curved Spacetime},\\
arXiv:0801.4676v1 [math-ph], to be published in Commun. Math. Phys.

\bibitem{Sanders}
K. Sanders, \emph{Aspects of locally covariant quantum field
theory}, PhD thesis, York (July 2008), also available as
arXiv:0809.4828v1 [math-ph]

\bibitem{Sanders2}
K. Sanders, in preparation

\bibitem{Schaefer}
H.H. Schaefer, \emph{Topological vector spaces},
Macmillan, New York (1966)
 
\bibitem{Strohmaier+}
A. Strohmaier, R. Verch and M. Wollenberg, \emph{Microlocal analysis
of quantum fields on curved space-times: analytic wavefront sets and
Reeh-Schlieder theorems},
J. Math. Phys. \textbf{43} (2002), 5514--5530

\bibitem{Verch}
R. Verch, \emph{A spin-statistics theorem for quantum fields on curved
spacetime manifolds in a generally covariant framework},
Commun. Math. Phys. \textbf{223}  (2001),  261--288

\bibitem{Wald}
R.M. Wald, \emph{Quantum field theory in curved spacetime and black
hole thermodynamics},
The University of Chicago Press, Chicago and London (1994)
\end{document}